\documentclass[10pt,twocolumn]{revtex4-1}

\usepackage{amsmath}
\usepackage{amssymb}
\usepackage{verbatim}
\usepackage{graphicx}
\usepackage{subfigure}
\usepackage{dcolumn}
\usepackage{multirow}
\usepackage{setspace}

\begin{document}
\title{{\Large Multi-compartment Linear Noise Approximation}}

\author{{\large Joseph~D. Challenger}}
\email{joseph.challenger@postgrad.manchester.ac.uk}
\affiliation{Theoretical Physics Division, School of Physics and
Astronomy, The University of Manchester, Manchester M13 9PL, UK}
\author{{\large Alan~J. McKane}}
\affiliation{Theoretical Physics Division, School of Physics and
Astronomy, The University of Manchester, Manchester M13 9PL, UK}
\author{{\large J\"{u}rgen Pahle}}
\affiliation{CICADA, School of Mathematics, The University of Manchester, Manchester Interdisciplinary Biocentre,
131 Princess St., Manchester, M1 7DN, UK}

\begin{abstract}
The ability to quantify stochastic fluctuations present in biochemical and other systems is becoming increasing important. Analytical descriptions
of these fluctuations are attractive, as stochastic simulations are computationally expensive.
Building on previous work, the linear noise approximation is developed for biochemical models with many compartments e.g.~cells.
The procedure is then implemented in the software package COPASI.
This technique is illustrated with two simple examples and is then applied to a more realistic biochemical model. Expressions for the noise, given in 
the form of covariances matrices, are presented.
\end{abstract} 

\pacs{87.18.Tt, 05.40.-a, 87.18.Vf}

\maketitle

\section{Introduction}
\label{sec:intro}
When modeling chemical or biochemical reactions, most of the time 
the particulate nature of the constituents is ignored, and their 
concentrations are treated as deterministic and continuous variables. This is the standard
thermodynamic approach to reaction kinetics. However there are a number of
situations where this approach is not valid \cite{rao_02,mcadams_99,mcadams_97,ullah_11}. These situations usually arise
in the case of biochemical reactions in a cellular environment. 
Here, there may only be hundreds, or even fewer, of molecules of one particular type and the
individual nature of the molecules cannot be disregarded.

It might be thought that, since statistical fluctuations involving $N$ 
particles typically have an effect of size $1/\sqrt{N}$ \cite{kampen_81}, 
the usual thermodynamical description would be approximately valid, with errors of
less than a few per cent for systems of this kind. This is not the case, 
as many simulations of biochemical reactions, and also experiments, have shown e.g. \cite{kummer_05,eldar_10}. There are cases where a 
stationary state with constant concentrations is expected from an analysis of 
the deterministic rate equations, but simulations of the stochastic model show sustained cycles and 
other complex oscillatory behavior \cite{mckane_05,mckane_07}. So, far from being only slightly 
quantitatively incorrect, the rate equations can give qualitatively wrong 
results.

There are two principal methods of investigating the effects of statistical 
fluctuations on reaction kinetics. The first is computational. When the 
reactions are formulated as a chemical master equation there are, formally 
exact, simulation algorithms which theoretically allow for a complete understanding of the
nature of the resulting behavior \cite{gillespie_76, pahle_09}. However they are time
consuming, and for many realistic biochemical reactions
prohibitive in terms of computational time. Fortunately a second,
analytic but approximate, approach is available which is usually accurate and easy to use. This is
the linear noise approximation (LNA) where stochastic deviations from the 
deterministic description are kept to linear order only. The spectrum of 
these fluctuations can then be calculated analytically \cite{mckane_05}. 

Both of these approaches have been systematically formulated in the last few 
years. Stochastic simulations are one important feature of the freely available software {COPASI} 
\cite{hoops_06,copasi} which gives researchers interested in modeling and simulation of 
biochemical networks easy access to a variety of different analysis and simulation methods.
The LNA 
may be formally derived from the chemical master equation by using the 
system-size expansion of van Kampen \cite{kampen_81}, and several new insights have 
been obtained in this way. In particular, the reason why fluctuations are 
still important when $N$ is relatively large can be understood as an 
amplification phenomenon; a resonance effect means that the size of the 
fluctuations is of the order of $C/\sqrt{N}$, where the constant $C$ is 
not of order one, but of order of $10^2$ or $10^3$ or even larger \cite{mckane_05,mckane_07}. 
Therefore fluctuations may be significant even when $N$ is of the
order of a million.

In this paper we further develop the theory of the LNA, discuss it in the 
context of COPASI, and apply the results to specific examples of biochemical 
reaction networks. In a previous paper \cite{pahle_11}, we showed how to incorporate the LNA 
into the COPASI program. There we combined the LNA with optimization or parameter scanning in a closed loop to study the correlation structure of biochemical systems and illustrated the use of this new strategy on several examples. However 
that study was restricted in an important way. In many situations molecules
react in a region with a given volume (denoted as a compartment), but may 
also be able to move from one compartment to another, possibly undergoing a chemical transformation on the way. The volumes of these 
compartments may differ, and these differences will have important effects 
on the effective concentration fluxes. In our previous work we avoided this 
complication by only investigating models with a single compartment. Biochemical models can, in principle, be reformulated to refer to only one compartment. 
However this is a serious restriction, as this procedure must be done `by hand' and is both time consuming and prone to error. 
Thus, in this paper, we show how to generalize the analysis to multi-compartment models, and so considerably extend the number 
of systems which can be investigated using our methodology.

The outline of the paper is as follows. In Sec.~\ref{sec:formalism} we 
introduce the formalism of the LNA when two compartments are present in the 
model and illustrate its use on a simple example. We then generalize this 
treatment in Sec.~\ref{sec:general} to many-compartment systems and discuss how the method 
can be used within COPASI, using a slightly more complicated model to 
illustrate this. An application to a biochemical model taken from the literature is given in Sec.~\ref{sec:applications} to show the power of the method. We 
conclude in Sec.~V with a discussion of possible future avenues of 
investigation. The Appendices contain some mathematical details, as well as some specific details of the models 
analyzed in the main text.

\section{Formalism}
\label{sec:formalism}
When discussing the LNA in detail, we shall use the same formalism as in our previous paper \cite{pahle_11}. The general biochemical model,
with $\hat{K}$ species  $Y_1,\ldots,Y_{\hat{K}}$ and $M$ reactions, will be written as
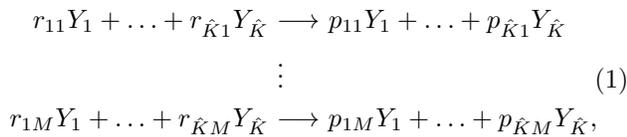
\begin{align}
r_{11}Y_1 + \ldots + r_{\hat{K}1}Y_{\hat{K}} \ &{\longrightarrow}
\ p_{11}Y_1 + \ldots +
p_{\hat{K}1}Y_{\hat{K}} \nonumber \\ &\vdots\\ r_{1M}Y_1 + \ldots +
r_{\hat{K}M}Y_{\hat{K}} \ &{\longrightarrow} \ p_{1M}Y_1 + \ldots
+ p_{\hat{K}M}Y_{\hat{K}}, \nonumber
\label{gen_chem}
\end{align}
where the $r_{i\mu}$ and $p_{i\mu}\ (i=1,\ldots,\hat{K};
\mu=1,\ldots,M)$ are the numbers of molecules of reactants
and products involved in each reaction. This may also be written as
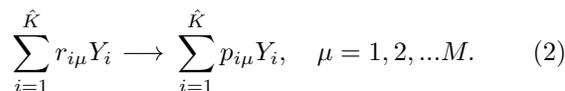
\begin{equation}
\sum_{i=1}^{\hat{K}} r_{i\mu}Y_i \ {\longrightarrow}
\ \sum_{i=1}^{\hat{K}} p_{i\mu} Y_i, \ \ \ \mu=1,2,...M.
\label{gen_reaction} 
\end{equation}
The elements of the
stoichiometric matrix, $\nu_{i\mu} \equiv p_{i\mu}-r_{i\mu}$, describe
how many molecules of species $Y_i$ are gained or lost due to
reaction $\mu$. Biochemical systems frequently contain conservation relations. That is, although there are $\hat{K}$ chemical species present
in the system, they may not all be able to vary independently. We denote the number of conservation 
relations by $\Lambda$ and the number of independent species by $K$=$\hat{K}-\Lambda$. As in \cite{pahle_11}, we work with a system of dimension 
$K$: information about the remaining species may be obtained from the conservation equations. 
When describing the biochemical model stochastically, we begin by defining a chemical master equation, where the state of the system is described
by $\boldsymbol{n}=(n_1,\ldots,n_K)$, the vector of particle numbers for species $Y_1,\ldots,Y_K$. 
This equation describes the time
evolution of $P(\boldsymbol{n},t)$: the probability for the system to be found in the state $\boldsymbol{n}$ at time $t$. It has the form \cite{gardiner_04}
\begin{equation}
\begin{split}
\frac{\mathrm{d}P(\boldsymbol{n},t)}{\mathrm{d}t} = \sum^{M}_{\mu=1}
 T_{\mu}(\boldsymbol{n}|\boldsymbol{n}-\boldsymbol{\nu}_{\mu})
P(\boldsymbol{n}-\boldsymbol{\nu}_{\mu},t) \\
-T_{\mu}(\boldsymbol{n}+\boldsymbol{\nu}_{\mu}|\boldsymbol{n})
P(\boldsymbol{n},t),
\end{split}
\label{master_eqn_1}
\end{equation}
where $T_{\mu}(\boldsymbol{n}'|\boldsymbol{n})$ is the transition rate from state $\boldsymbol{n}$ to $\boldsymbol{n'}$ due to reaction $\mu$.
The new state $\boldsymbol{n'}$ is defined by the stoichiometry of the reaction, $\boldsymbol{\nu}_{\mu}=(\nu_{1\mu},\ldots,\nu_{K\mu})$.
To illustrate how the method applies to systems with more than one compartment we shall look at a simple system involving the diffusion
of molecules between two compartments. Molecules in the first compartment, of volume $V_1$, are labeled $D$, 
molecules in the second compartment, volume $V_2$, are labeled $E$.
The system is summarized in Table~\ref{tab:reaction1}, where $n_1$ is the number of $D$ molecules, and $n_2$ is the number of $E$ molecules. 
There is an extra complication to consider when defining the transition rates for a system 
with many compartments. That is, how do the transition rates of compartment-crossing reactions scale with the volumes of the compartments? 
The rates of transitions from one compartment to another one should be dependent on properties of the contact surface between the compartments. 
These properties, such as the size of the diffusion area or the number of channels, will generally not scale with 
any compartment volume in practice. Rather, this relation will be different for different reaction systems, and will 
depend on the geometry of the overall system, and on how the compartments are connected.
The transition rates for both reactions considered here are proportional to the concentration of the substrate, and scale 
with the volume of the first compartment. In this paper, we shall just consider transition rates that scale linearly with volume in this way. 
We shall return to this issue in Sec.~\ref{sec:discussion}.
\begin{table}[!ht]
\caption{A two-compartment reaction system}
\begin{flushleft}
A simple reaction system showing diffusion between two compartments.
\end{flushleft}  
\begin{tabular}{|c|c|c|c|}
\hline
            & Reaction        & Stoichiometry	& Transition Rate           \\
\hline
1 & $D \longrightarrow E $ &  $\boldsymbol{\nu}_1=(-1,1)$ & $T_1(\boldsymbol{n}+\boldsymbol{\nu}_1|\boldsymbol{n})=kn_1 $ \\

2 & $E \longrightarrow D $   & $ \boldsymbol{\nu}_2=(1,-1) $ & $T_2(\boldsymbol{n}+\boldsymbol{\nu}_2|\boldsymbol{n})=k\frac{V_1}{V_2}n_2 $ \\

\hline
\end{tabular}
\label{tab:reaction1}
\end{table}
There is a conserved quantity present in the system: the total number of molecules, $u$, is unaffected by each reaction i.e.
$n_1+n_2=u$. Therefore, the system has
only one independent variable, so we can describe the system by either the number of $D$ or $E$ molecules present. 
Here we choose to use $D$. We rewrite the transition rate for the second reaction as $T_2=k(V_1/V_2)(u-n_1)$, 
using the conservation equation. The contributions from each reaction to Eq.~\eqref{master_eqn_1} may be written as
\begin{align}
\textrm{Reaction 1:} \,\,\, &(\mathbb{E}^{+1} -1)[kn_1 P(\boldsymbol{n},t)],\nonumber \\
\textrm{Reaction 2:} \,\,\, &(\mathbb{E}^{-1} -1)[k\frac{V_1}{V_2}(u-n_1) P(\boldsymbol{n},t)],\nonumber \\
\end{align}
where the $\mathbb{E}$ are step operators, e.g. $\mathbb{E}^{-1}f(n_1)=f(n_1-1)$. To perform the van Kampen
expansion \cite{kampen_81}, we split $n_1$ into a deterministic part, the macroscopic concentration, $x$, and a stochastic part, $\xi$:
\begin{equation}
 \frac{n_1}{V_1}= x+\frac{\xi}{\sqrt{V_1}}.
\label{kampen}
\end{equation}
After the change of variables given in Eq.~\eqref{kampen}, the step operator may be written as
\begin{equation}
 \mathbb{E}^{\pm 1}=1\pm\frac{1}{\sqrt{V_1}}\frac{\partial}{\partial \xi}+\frac{1}{2V_1}\frac{\partial^2}{\partial \xi^2}\ldots \ \ .
\label{step_op}
\end{equation}
The master equation is then rewritten in terms of the new variables and terms of the same order are equated. The first order terms recover
the macroscopic description of the system. That is, an ODE for the chemical concentration $x$
\begin{align}
 \frac{\mathrm{d}x}{\mathrm{d}t}&=k[\alpha^2(\tilde{u}-x)-x],\nonumber \\
\label{ode_d_e}
\end{align}
where $\alpha^2=V_1/V_2$ and $\tilde{u}=\textrm{lim}_{V_1 \to \infty}u/V_1$. Thus, the volume ratio becomes an extra parameter of the problem.
The macroscopic equation has a stable fixed point \cite{strogatz}, which we shall denote by $x=x^*$. 
The second order terms define a Fokker-Planck equation \cite{kampen_81} for $\Pi(\xi,t)$, the probability 
distribution of the random variable $\xi$,
\begin{equation}
\frac{\partial\Pi}{\partial t}=
-\frac{\partial}{\partial \xi}(A\xi\Pi)
+\frac{1}{2}B
\frac{\partial^{2}\Pi}{\partial \xi^2}.
\label{FPE}
\end{equation}
This equation is fully characterized by $A$ and $B$, which are functions of the macroscopic concentration of the chemical species, $x(t)$. 
In this example it is found that they are
\begin{align}
 A= -k(\alpha^2+1),\nonumber \\
 B= kx+k\alpha^2(\tilde{u}-x).\nonumber \\
\end{align}
The general form of $A$ and $B$ will be discussed in Sec.~\ref{sec:general}. 
For models with $K>1$, the Fokker-Planck equation will be multi-variate, and $A$ and $B$ will be square matrices of dimension $K$. 
$A$ and $B$ are often called the drift and diffusion matrices, respectively. 
When they are evaluated at the fixed point, $x=x^*$, $A$ and $B$ become constant and
$\Pi$ describes the fluctuations around the macroscopic steady state.
From the Fokker-Planck equation, information about the moments may be obtained. 
Of particular interest in this paper are the covariances,  $\Xi_{ij}=\langle( \xi_i -\langle \xi_i \rangle )(\xi_j-\langle \xi_j \rangle)\rangle$.
The matrix of covariances satisfies the following matrix equation \cite{kampen_81} 
\begin{equation}
 A\Xi+\Xi A^T+ B=0, 
\label{lyap1}
\end{equation}
which is known as the Lyapunov equation \cite{horn_91}. 
For our one-dimensional example, $\Xi$ is just the variance of the random variable $\xi$ 
i.e. $\Xi=\langle( \xi -\langle \xi \rangle )(\xi-\langle \xi \rangle)\rangle$. 
As discussed in our previous work, we are more interested in the variance in terms of molecular
numbers $n_1$, rather than $\xi$. That is, the desired form of the variance, $C$, in this example is 
$C=\langle( n_1 -\langle n_1 \rangle )(n_1-\langle n_1 \rangle)\rangle$. From our previous paper \cite{pahle_11} (see also \cite{elf_03b}) it is possible to link
the two quantities by $C=V_1 \Xi$. The general form of the relationship between $C$ and $\Xi$ in a model with many compartments
will be discussed in Appendix~\ref{sec:append_a}.
For one-dimensional problems, Eq.~\eqref{lyap1} can be trivially rearranged to solve for $\Xi$, the variance of the fluctuations around the 
steady state. We can compare this result from theory with that obtained 
via simulation, using the Gillespie algorithm \cite{gillespie_76}, as implemented in COPASI. The value for the variance found by
using the LNA was 360, in units of particle numbers squared, compared to 360.0 (to one decimal place) obtained from simulation, averaged over 5000 time series. 
The numerical values used for the parameters in this example are as follows: $V_1=0.3$, $V_2=0.2$ (both in picolitres), 
$k=0.2$ and the total number of molecules in the system was 1500. 

\section{General Case}
\label{sec:general}
As mentioned in the previous section, the $A$ and $B$ matrices are functions of the macroscopic concentrations. But the transition rates that
define the master equation depend on the molecular populations, which are discrete quantities. So, to define the general form of $A$ and $B$ we must define
a macroscopic quantity $F_{\mu}(\boldsymbol{x})$ that corresponds to the transition rate 
$T_{\mu}(\boldsymbol{n}+\boldsymbol{\nu}_{\mu}|\boldsymbol{n})$ for reaction $\mu$. 
To do this, we simply make the replacement $n_i/V^{(i)}\rightarrow x_i$, since $\lim_{V^{(i)} \to \infty}\langle n_i \rangle/V^{(i)}$ becomes
equal to $x_i$ in the thermodynamic limit. Here, $V^{(i)}$ denotes the volume of the compartment within which species $i$ is located. In this paper all 
volumes written with bracketed superscripts will be defined in this way: volumes with subscripts define the actual volume of
a compartment. The general forms of $A$ and $B$ look very similar to those given in \cite{pahle_11}, except for extra volume factors, which are picked up 
from the step operators, as shown in Eq.~\eqref{step_op}. In one compartment models these volume factors cancel. 
The general form for $A$ and $B$ for multi-compartment models are 
\begin{align}
A_{ij}(\boldsymbol{x})
&= \sum_{\mu=1}^M \frac{\nu_{i\mu}}{\sqrt{V^{(i)}V^{(j)}}}\frac{\partial F_\mu(\boldsymbol{x})}{\partial x_j} \ \ i,j=1,\ldots,K, \nonumber \\
 B_{ij}(\boldsymbol{x})
&=\sum_{\mu=1}^M \frac{\nu_{i\mu}\nu_{j\mu}}{\sqrt{V^{(i)}V^{(j)}}}F_{\mu}(\boldsymbol{x}) \ \ i,j=1,\ldots,K.  \nonumber \\ 
\label{gen_a_b}
\end{align}
It is convenient to define the above matrices in terms of the $F_{\mu}(\boldsymbol{x})$, since these are quantities which COPASI already 
calculates. In COPASI they are called `particle fluxes', and are calculated as part of the Steady State Task. 

In one compartment models the entries of the matrix $A$ are found to be identical to the entries of the
Jacobian of the macroscopic system evaluated at the fixed point. If the general form of the ODEs is taken to be 
$\mathrm{d}x_i/\mathrm{d}t=g_i(\boldsymbol{x})$, then $A_{ij}=\partial g_i/\partial x_j$. This is not the case in models with many
compartments. However, the correct form of $A$ for many compartment models
can be found by applying a similarity transformation to the Jacobian, which from now on we will call $\tilde{A}$. The relationship is found to be
\begin{equation}
 A = S\tilde{A}S^{-1}, \ \ S=\textrm{diag}(\sqrt{V^{(1)}},\sqrt{V^{(2)}},\ldots,\sqrt{V^{(K)}}).
\label{eq12}
\end{equation}
As COPASI is able to calculate the Jacobian, we will utilize this calculation, instead of
performing an extra one. However, there is an additional complication here, since COPASI uses ODEs for the expectation of the number of
molecules of the chemical species, rather than their concentrations. The Jacobian calculated from the former, which we will denote $\hat{A}$, 
is not identical to the one calculated from the latter, $\tilde{A}$. 
However, these two matrices are similar, and can be easily related to each other (see Appendix~\ref{sec:append_a} for details):
\begin{equation}
 \hat{A}= S_2 \tilde{A} S_2^{-1}, \ \  S_2=\textrm{diag}(V^{(1)},V^{(2)},\ldots,V^{(K)}).
\label{a_hat_tilde}
\end{equation}
We can use this relation, along with the relation linking $\tilde{A}$ with $A$, the desired matrix, to find the relation between $\hat{A}$ and
$A$. It is $\hat{A}=SAS^{-1}$.

We want to use these matrices to define a Lyapunov equation which can be solved to yield the covariances. One course of action would be to convert
$\hat{A}$ to the desired form, $A$, and use this to solve Eq.~\eqref{lyap1} for $\Xi$ and convert to $C$, the covariances in terms of 
particle numbers, using Eq.~\eqref{c_xi}. It is slightly more straightforward, however, to define an equivalent Lyapunov equation, 
involving $\hat{A}$ and $C$,
\begin{equation}
 \hat{A}C+C\hat{A}^T +\hat{B}=0,
\label{lyap_new}
\end{equation}
where, in order for the above equation to be equivalent to Eq.~\eqref{lyap1}, we make the identification $\hat{B}=SBS$. This is equivalent to defining $B$ in
Eq.~\eqref{gen_a_b} without the square rooted volume factors in the denominator. This is the form displayed in COPASI. 
COPASI then solves this equation for $C$, using the Bartels-Stewart algorithm \cite{bartels_72}. 
The equivalence of the two Lyapunov equations is proved in Appendix~\ref{sec:append_a}.

\begin{table}[!ht]
\caption{The three-compartment reaction system}
\begin{tabular}{|c|c|c|}
\hline
            & Reaction        & Particle Flux           \\
\hline
1 & $ G_6 \longrightarrow G_1 $ &  $F_1(\boldsymbol{x})=k_1x_6V_3$   \\

2 & $2G_1 \longrightarrow 3G_2 + 4G_3 $   & $F_2(\boldsymbol{x})=k_2x_1^2V_1  $   \\

3 & $ G_2 \longrightarrow ;G_5$ &  $F_3(\boldsymbol{x})=k_3x_2x_5V_2  $    \\

4 & $ G_3 \longrightarrow $ &  $F_4(\boldsymbol{x})=k_4x_3V_3  $   \\

5 & $ G_4 \longrightarrow G_5 $ &  $F_5(\boldsymbol{x})=k_5x_4V_1 $   \\

6 & $ G_5 \longrightarrow G_4 $ &  $F_6(\boldsymbol{x})=k_6x_5V_1  $   \\

7 & $ G_1 \longrightarrow G_6 $ &  $F_7(\boldsymbol{x})=k_7x_1V_3  $   \\

8 & $  \longrightarrow G_1 $ &  $F_8(\boldsymbol{x})=k_8V_1 $    \\

\hline
\end{tabular}

\label{tab:reaction2}
\end{table}

We end this section with another example, this time with three compartments instead of two, to illustrate the natural extension of 
the method to an arbitrary number of compartments. 
This three compartment model has six species. Species $G_1$ and $G_4$ are located in compartment of volume
$V_1$, $G_2$, $G_3$ and $G_5$ in compartment of volume $V_2$ and $G_6$ in compartment with volume $V_3$. The number of $G_1$ molecules is denoted by $n_1$,
the number of $G_2$ molecules by $n_2$, and so on.
The reactions are described in Table~\ref{tab:reaction2}, along with their particle fluxes.

In reaction 3, the rate of degradation of species $G_2$ is now modified by the concentration of $G_5$ within the compartment. We notice that
species $G_4$ and $G_5$ are linearly dependent on each other, and we choose to eliminate $G_5$. The matrices $A$ and $B$ are given in 
Appendix~\ref{sec:append_b}. COPASI solves the Lyapunov 
equation for the 5 dimensional `reduced' system, then recovers the full 6 dimensional covariance matrix using the conservation 
relations, see \cite{reder_88,pahle_11} for details.

Table~\ref{tab:c6} shows the numerical values obtained for the covariances. The parameter values were chosen to be $k_1=0.1\textrm{s}^{-1}$, 
$k_2=0.02\textrm{pl}\#^{-1}\textrm{s}^{-1}$, $k_3=0.1\textrm{pl}\#^{-1}\textrm{s}^{-1}$, $k_4=2\textrm{s}^{-1}$, 
$k_5=0.1\textrm{s}^{-1}$, $k_6=0.1\textrm{s}^{-1}$, $k_7=0.1\textrm{s}^{-1}$ and $k_8=50\#\textrm{pl}^{-1}\textrm{s}^{-1}$, where $\#$ denotes particle numbers. 
The compartment volumes are $V_1=8$, $V_2=100$ and $V_3=72$ (in picolitres). Results for 
species $G_5$ are not given, but may be found from conservation considerations. The conservation relation for
this system is $n_4+n_5=1300$ molecules. 

\begin{table}[!h]
 \caption{Covariances for the three-compartment system }
\begin{flushleft}
The covariances of the fluctuations around the steady state, in units of particle numbers squared. 
Results obtained from the LNA are compared with those from simulation (in brackets), via the Gillespie algorithm. 
10000 time series, each of length 10000 seconds with 5000 samples, were generated. 
\end{flushleft}
\begin{tabular}{|c|c|c|c|c|c|}
\hline
  & $G_1$ & $G_2$& $G_3$ & $G_4$ & $G_6$ \\
\hline

$G_1$  & 229 (229) & -48 (-48) & -55 (-55) & 0 (0) & -17 (-17) \\
$G_2$ & -48 (-48) & 841 (841) & 592 (592) & 34 (34) & -87 (-88)  \\
$G_3$  & -55 (-55) & 592 (592) & 846 (846) & 0 (0) & -68 (-68) \\
$G_4$ & 0 (0) & 34 (34) & 0 (0) & 89 (89) & 0 (0)\\
$G_6$ & -17 (-17) & -87 (-88) & -68 (-68) & 0 (0) & 2396 (2393) \\
\hline
\end{tabular}
\label{tab:c6}
\end{table}

\section{A Biochemical Application}
\label{sec:applications}
In this section we look at a model due to Kongas and van Beek \cite{kongas_07} that studies the role of creatine kinase in the heart by 
examining energy metabolism in cardiac muscle. It is a two-compartment system, with a cytoplasm 
and an intermembrane space. We shall label the volumes of these compartments as $V_c$ and $V_i$ respectively. In the original article the model is studied 
deterministically. Here, we will
reduce the model in size, without changing the volume ratio used in the article, to study it stochastically. The model is described schematically in Figure 1 
of \cite{kongas_07}. It involves 5 chemical species, $ADP$, $ATP$, creatine ($Cr$), phosphocreatine ($PCr$) and inorganic phosphate ($Pi$). 
All of these metabolites are present in both compartments, so we have 10 variables. We use a subscript, $i$, to denote the species in compartment $V_i$. 
The reactions are as follows: 
 \begin{align}
  ADP_i + Pi_i &\rightleftharpoons ATP_i \nonumber \\
  ATP_i + Cr_i &\rightleftharpoons ADP_i +PCr_i \nonumber \\
  ATP + Cr &\rightleftharpoons PCr + ADP \nonumber \\
  ATP &\rightarrow ADP + Pi \nonumber \\
  Pi_i &\rightleftharpoons Pi \nonumber \\
  Cr_i &\rightleftharpoons Cr \nonumber \\
  ADP_i &\rightleftharpoons ADP \nonumber \\
  PCr_i &\rightleftharpoons PCr \nonumber \\
  ATP_i &\rightleftharpoons ATP.  \nonumber \\
\label{CK_reactions}
 \end{align}

A SBML (Systems Biology Markup Language \cite{hucka_03}) implementation of the model is available from the BioModels Database \cite{biomodels_10,biomodels_web}. 
This file, which can be downloaded and then read by COPASI, corrects an error in the rate equations given in \cite{kongas_07}. 
As for our previous models, not all of the 10 species can vary independently, 
as there are three conservation relations present. Hence, COPASI reduces the dimensionality
of the model to 7. The conservation relations, in terms of molecule numbers, are 
\begin{eqnarray}
ATP_i+ADP_i+ATP+ADP=C_1, \nonumber \\  
P_i-ADP_i+PCr+Pi_i-ADP+PCr_i=C_2, \nonumber \\
Cr+Cr_i+PCr+PCr_i=C_3, \nonumber \\
\label{CK_conservation}
\end{eqnarray}
where $C_1$, $C_2$ and $C_3$ are integer constants. We chose the values $C_1=35000,C_2=1204,C_3=11743$ and did not alter the reaction 
parameters given in \cite{kongas_07}.
The values for $C_1$, $C_2$ and $C_3$ were chosen to speed up the numerical simulations of the system 
(which are extremely slow) by reducing the overall number of molecules in the system, whilst ensuring that each species did not get too close to the zero
particle boundary. The deterministic model of the system has a unique steady-state, which is described in Table~\ref{tab:reaction3}. 
We calculated the covariances of the fluctuations around the steady state using the LNA Task in COPASI. Table~\ref{tab:ck} shows these results
and compares the values with those obtained from numerical simulation. Only results for 7 of the 10 species present in the model are shown: 
the covariances for the other species may be obtained from the conservation equations.

\begin{table}[!ht]
\caption{The creatine kinase model}
\begin{flushleft}
Description of the biochemical system at the steady-state. Steady-state values are given in terms of particle numbers, and are rounded to the nearest integer.
\end{flushleft}
\begin{tabular}{|c|c|c|}
\hline
Compartment         & Species        & Steady-State Value          \\
\hline
 & $ADP$ & 17490   \\

 & $ATP$    & 10   \\

$V_c$ & $Cr$ &  5859    \\

 & $PCr$  &  13  \\

 & $Pi$  &  16501   \\
\hline
 & $ADP_i$  & 11819  \\

 & $ATP_i$   &  5681 \\

$V_i$ & $Cr_i$ &  5857 \\

 & $PCr_i$ & 15   \\

 & $Pi_i$ & 13984  \\
\hline
\end{tabular}

\label{tab:reaction3}
\end{table}

\begin{table}[h]
 \caption{Covariances for the creatine kinase model}
\begin{flushleft}
 Values for the covariances of the fluctuations around the steady state, in units of particle numbers squared. 
Results obtained from the LNA are compared with those found from
numerical simulation (in brackets), via the Gillespie algorithm. 4000 time series, each of length 1000 seconds with 10000 samples, were generated. 
The typical standard deviations associated with the simulation results are less than 1. The largest, for the variances of $ADP_i$ and $ATP_i$, are 1.8.
\end{flushleft}

\begin{tabular}{c|c|c|c|c|c|c|c|}
\cline{2-8}
  & $ADP_i$ & $ATP$ & $Cr_i$ & $PCr$ & $Pi_i$ & $ATP_i$ & $Cr$ \\
\hline

\multicolumn{1}{|c|}{\multirow{2}{*}{$ADP_i$}}  & 8407  & -4  & 1  & 0  & 1512 &-2252 &-1 \\
\multicolumn{1}{|c|}{}	  &(8406) &  (-4) & (1) & (0) &  (1511) &(-2252)	&(-1) \\
\hline
\multicolumn{1}{|c|}{\multirow{2}{*}{$ATP$}} & -4 & 10  & 0 & 0 & -5 	&-2	& 0 \\
\multicolumn{1}{|c|}{}	      &(-4)  & (10)&(0) &(0) & (-5)	&(-2)&(0)  \\
\hline
\multicolumn{1}{|c|}{\multirow{2}{*}{$Cr_i$}}  & 1 & 0 & 2936  & -6  & 8 &-1	&-2922	\\
\multicolumn{1}{|c|}{}		& (1) & (0) & (2936) & (-6) &  (7) &(-1)&(-2922)	\\
\hline
\multicolumn{1}{|c|}{\multirow{2}{*}{$PCr$}} & 0  & 0  & -6  & 13  & -7  & 0	& -6	\\
\multicolumn{1}{|c|}{}	&  (0) & (0) & (-6) & (13) & (-7) &	(0)& (-6)	\\
\hline
\multicolumn{1}{|c|}{\multirow{2}{*}{$Pi_i$}} & 1512 & -5  & 8 & -7 & 9467  & -2732	& 7	\\
\multicolumn{1}{|c|}{}		&  (1511) &  (-5) &  (7) &  (-7) & (9469) &(-2731)	&(7)	\\
\hline
\multicolumn{1}{|c|}{\multirow{2}{*}{$ATP_i$}} &-2252  & -2  & -1  & 0  & -2732	& 4846	& 0\\
\multicolumn{1}{|c|}{}		& (-2252) & (-2) & (-1) & (0) &  (-2731)&(4846)&(1)\\
\hline
\multicolumn{1}{|c|}{\multirow{2}{*}{$Cr$}} & -1  & 0 & -2922 & -6  & 7 	& 0	&	2936 \\
\multicolumn{1}{|c|}{}	      & (-1) &  (0) & (-2922) & (-6) & (7)	&(1)	&(2936) \\
\hline
\end{tabular}
\label{tab:ck}
\end{table}

\section{Discussion}
\label{sec:discussion}

In our previous work \cite{pahle_11} we automated the LNA procedure for one
compartment models into the software package COPASI. This allowed the 
fluctuations around a steady-state to be found without employing
time-consuming algebra. It also means that fluctuation
analysis may be calculated in a closed-loop with other tasks available in
COPASI, e.g. optimization.  In this paper we extended this
framework to include multi-compartment models. The accuracy of the
method was demonstrated with simple examples in Sec.~\ref{sec:formalism} and \ref{sec:general}. In
Sec.~\ref{sec:applications}, a more realistic model was considered. A general formalism for the LNA 
for multi-compartment models has not been proposed before. However, Ullah and Wolkenhauer \cite{ullah_09,ullah_11} 
have used the idea of relative concentrations for models which do not have a single system-size parameter for all 
species. Although the motivation for this was not to study multi-compartment models, the formalism is similar to 
the one outlined here. 

Some issues remain which have not yet been resolved. 
Reaction kinetics for cross-compartment reactions need to be
carefully considered. For convenience, in many published models cross-compartmental reactions
are often defined so that they scale with volume in the same way as
reactions in the `bulk' of the cell: this is the approach we have taken in this paper. In reality this will not 
always be the case, as mentioned in Sec.~\ref{sec:formalism}. The LNA will work with any particular scaling that is universal for the system.
However, a consistent methodology for analyzing fluctuations in systems with a mixture of scalings present 
(e.g. some reactions scale with a cell's surface area and some with its volume) has not yet been proposed. 
Note that when describing the reaction systems, no
mention of geometry of the compartments has been made. Any geometrical
considerations should be reflected in the specific form of the transition rates
chosen by the user.

Care should be taken when using the LNA in certain circumstances. 
As we mentioned in \cite{pahle_11}, the
Gaussian assumption of the noise breaks down if the system is close to
a boundary e.g. zero molecules. 
In addition to this, the LNA for multi-compartment models is technically only valid
when the compartment volumes are of comparable order. This is because terms of
order e.g. $(1/\sqrt{V_1})$ and 
$(1/\sqrt{V_2})$ are equated when performing the expansion of
the master equation. It is possible to go beyond the LNA by retaining terms of higher orders in the expansion. This can be done in a systematic 
way \cite{kampen_81,grima_10,ullah_11}. In most cases, the additional terms yield only very small corrections, although there do exist 
cases in which larger deviations have been found \cite{goutsias_07,ferm_08,grima_10,thomas_12}. In particular, a shift in the mean can be detected when the compartment volume(s) are small. That is, the mean value of the stochastic process no longer coincides with the prediction from the deterministic model.  This is demonstrated in \textit{e.g.} \cite{thomas_12}, for systems with volume of order $1\textrm{fl}$ or smaller. The shift of the mean was found analytically using the higher order terms in the expansion method used here. We remark that the systems used in that work contain fewer  molecules than those used here, so these higher order effects become more significant. Moment closure techniques have also been used to explain the discrepancy between the average behaviours of the deterministic and stochastic systems \cite{goutsias_07}. However, we do not favour such techniques, as they are not systematic. 
In all cases, stochastic simulation can be used to test the analytic approximations. These simulations can also be carried out within COPASI. One extension of the work performed here would be the calculation of the covariance matrix for the fluctuations from both numerical simulation and the LNA, carried out and displayed together for easy comparison. 

The multi-compartment approach taken here should not be confused with the treatment of a diffusion process \cite{kampen_81}. For a diffusion 
process, all of space is divided up into a large number of patches of equal size in order to study the spatial dynamics of the system, whereas each of 
our compartments is taken to be well-mixed and, unlike the patches, has a definite physical interpretation.

Studying biochemical models \textit{in silico} is an important method
by which to study biochemical systems. As models become more
complicated (and therefore more realistic), more and more models will
include many compartments. Therefore it is important to describe how the LNA can be used to study multi-compartment models. 
It is possible to transform multi-compartment models to a one compartment model, 
(as we did in \cite{pahle_11}) but this must be done `by hand' and is time consuming. With the framework described in this paper, 
models may be studied in their original form, using our extended implementation of the LNA in COPASI. We believe that this will be a
valuable tool in the study of stochastic effects in biochemical systems.

\section*{Acknowledgments}
\label{sec:acknowledgments}

JP thanks the UK's BBSRC (grant BB/F018398/1). JDC thanks EPSRC for the award of a PhD studentship.

\appendix
\section{Mathematical details}
\label{sec:append_a}
\subsection{Covariance transformation}
Here we use a simple example to show how the two covariance matrices, $C$ and $\Xi$, are related. The number
of molecules of species 1 is $n_1$ and the number of molecules of species 2 $n_2$. Species 1 is located within a compartment of volume $V_1$, 
species 2 within a compartment of volume $V_2$. The change of variables given in Eq.~\eqref{kampen} leads to
\begin{equation}
\label{vk}
n_1 = V_1x_1 +\sqrt{V_1}\xi_1, \ \ 
n_2 = V_2x_2 +\sqrt{V_2}\xi_2.
\end{equation}
The quantity $C_{12}$ is defined as
\begin{equation}
 C_{12}=\langle( n_1 -\langle n_1\rangle )(n_2-\langle n_2 \rangle)\rangle.
\end{equation}
From Eq.~\eqref{vk} we can express the expectation values of $n_1$ and $n_2$ to be 
\begin{equation}
\label{mean}
\langle n_1 \rangle = V_1x_1, \ \ \langle n_2 \rangle = V_2x_2.
\end{equation}
Note that this is only true to the order up to which we are working, (see e.g. \cite{kampen_81,grima_10,ullah_11} for details). 
Using the three equations above, we find that
$C_{ij}=\sqrt{V_1V_2}\langle \xi_i\xi_j\rangle$, which is equal to $\sqrt{V_1V_2} \Xi_{ij} $, where $\Xi$
is the covariance matrix for the random variable \textbf{$\xi$}. 
In general, $C_{ij}$, the covariance between species $i$ and $j$ in terms of particle numbers can be written as
\begin{equation}
 C_{ij}=\sqrt{V^{(i)}V^{(j)}}\Xi_{ij}.
\end{equation}
This relationship can also be expressed as a matrix transformation. For a general system, with $K$ species, the relationship is
\begin{equation}
 C=S\Xi S, \ \ S=\textrm{diag}(\sqrt{V^{(1)}},\sqrt{V^{(2)}},\ldots,\sqrt{V^{(K)}}).
\label{c_xi}
\end{equation}

\subsection{Relation between $\hat{A}$ and $\tilde{A}$}
In Section \ref{sec:general} we discussed the relationship between matrix $A$, calculated from the van Kampen expansion, and $\hat{A}$, the
form of the Jacobian calculated by COPASI, using ODEs for the expectation of the particle numbers. Eq.~\eqref{a_hat_tilde} gives the 
relationship between $\hat{A}$ and $\tilde{A}$, the Jacobian calculated by using ODEs for the concentrations of the chemical species. We will
highlight the differences between these quantities with a simple example, just considering reactions 1 and 7 in Table~\ref{tab:reaction2}.
These reactions make the following contributions to the macroscopic rate equations,

\begin{align}
 \frac{\mathrm{d}x_1}{\mathrm{d}t}&=\frac{V_3}{V_1}(k_1x_6-k_7x_1),\nonumber \\
 \frac{\mathrm{d}x_6}{\mathrm{d}t}&=(k_7x_1-k_1x_6).\nonumber \\
 \label{ode_a_f}
\end{align}
From the above equations, the entries of $\tilde{A}$ are found to be
 \begin{align}
 \tilde{A}_{11}=-\beta^2k_7, \ \ &\tilde{A}_{16}=\beta^2k_1, \nonumber \\
 \tilde{A}_{61}=k_7, \ \ &\tilde{A}_{66}=-k_1, \nonumber \\
\label{a_tilde}
\end{align}
where $\beta^2=V_3/V_1$. We can rewrite Eq.~\eqref{ode_a_f} in terms of the $\langle n_1 \rangle$ and $\langle n_6 \rangle$ 
\begin{align}
\frac{\mathrm{d}\langle n_1 \rangle}{\mathrm{d}t}&=  k_1\langle n_6 \rangle-k_7\frac{V_3}{V_1} \langle n_1 \rangle, \nonumber\\
\frac{\mathrm{d}\langle n_6 \rangle}{\mathrm{d}t}&= k_7 \frac{V_3}{V_1} \langle n_1 \rangle-k_1\langle n_6 \rangle. \nonumber \\
\label{ode_exp}
\end{align}
For the equations above, the elements of $\hat{A}$ are
\begin{align}
 \hat{A}_{11}=-\beta^2k_7, \ \ &\hat{A}_{16}=k_1, \nonumber \\
 \hat{A}_{61}= \beta^2k_1, \ \ &\hat{A}_{66}=-k_1. \nonumber \\
\label{a_hat}
\end{align}
By considering how the $x_i$ vary compared to the $\langle n_i \rangle$ it is possible to find the following relation by inspection
\begin{equation}
 \hat{A}_{ij}=\frac{V^{(i)}}{V^{(j)}}\tilde{A}_{ij}.
\end{equation}
Again, this may be written as a matrix transformation. For a general system,
\begin{equation}
 \hat{A}= S_2 \tilde{A} S_2^{-1}, \ \  S_2=\textrm{diag}(V^{(1)},V^{(2)},\ldots,V^{(K)}).
\end{equation}
A similar relation may be found between $A$ and $\tilde{A}$
\begin{equation}
 A=S\tilde{A}S^{-1}, \ \  S=\textrm{diag}(\sqrt{V^{(1)}},\sqrt{V^{(2)}},\ldots,V^{(K)}).
\end{equation}
Putting all this together, we find a relation between $A$ and $\hat{A}$
\begin{equation}
 \hat{A}=SAS^{-1}.
\label{aahat}
\end{equation}

\subsection{The equivalence of the Lyapunov equations}
Here we will prove the equivalence of Eq.~\eqref{lyap1} and Eq.~\eqref{lyap_new}, the two Lyapunov equations defined in the paper.
We start by rewriting equation \eqref{lyap1} as
\begin{equation}
AS^{-1}S\Xi+\Xi SS^{-1}A^T+ B=0,
\label{equiv1}
\end{equation}
where $S$ has its usual form. Next we pre- and post-multiply by $S$:
\begin{equation}
SAS^{-1}S\Xi S+S\Xi SS^{-1}A^TS+ SBS=0.
\label{equiv2}
\end{equation}
Using Eq.~\eqref{c_xi}, Eq.~\eqref{aahat} and defining $\hat{B}=SBS$ we can reduce the above equation to
\begin{equation}
\hat{A}C+C\hat{A}^T +\hat{B}=0,
\label{equiv3}
\end{equation}
which recovers \eqref{lyap_new}, the form of the Lyapunov equation solved by COPASI.

\section{Model details}
\label{sec:append_b}
For the reaction system described in Table~\ref{tab:reaction2} the form of $A$ and $B$ are as below, where $\alpha^2=V_1/V_2$ and $\beta^2=V_3/V_1$. 
The species are ordered $G_1$, $G_2$, $G_3$, $G_4$ and $G_6$: species $G_5$ having been eliminated due to conservation. 
The concentrations $\boldsymbol{x}$ are evaluated at the fixed point values $\boldsymbol{x}=\boldsymbol{x^*}$. $A$ and $B$ are 
\begin{widetext}
\begin{equation}
 A =
\left( {\begin{array}{ccccc}
 -4k_2x_1-k_7\beta^2 & 0 & 0 & 0 & k_1\beta \\
  6\alpha k_2x_1 & -k_3\Gamma & 0 & \alpha^2 k_3x_2 & 0  \\
 8\alpha k_2x_1 & 0 & -k_4 & 0 & 0\\
0 & 0 & 0 & -(\alpha^2k_6 + k_5) & 0 \\
k_7\beta & 0 & 0 & 0 & -k_1 \\
 \end{array} } \right),
\label{a2}
\end{equation}

\begin{equation}
 B =
\left( {\begin{array}{ccccc}
 \beta^2(k_1x_6 +k_7x_1)+ 4k_2x_1^{2}+k_8 & -6k_2\alpha x_1^{2} & -8k_2 \alpha x_1^{2} & 0 & -\beta(k_1x_6+k_7x_1) \\
 -6k_2\alpha x_1^{2} & 9k_2\alpha^2 x_1^{2} +k_3x_2\Gamma & 12k_2\alpha^2x_1^{2} & 0 & 0 \\
 -8k_2 \alpha x_1^{2} &  12k_2\alpha^2 x_1^{2} & 16k_2\alpha^2 x_1^{2}+k_4 x_3 & 0 & 0 \\
0 & 0 & 0 & k_5x_4+k_6\Gamma & 0 \\
-k_1x_6\beta -k_7x_1\beta& 0 & 0 & 0 & k_1x_6+k_7x_1 \\
 \end{array} } \right),
\label{b2}
\end{equation}
where $\Gamma=\tilde{u}-\alpha^2x_4$.

\end{widetext}

\end{document}